\documentclass[twocolumn,showpacs,superscriptaddress]{revtex4}

\usepackage{graphicx}%
\usepackage{dcolumn}
\usepackage{amsmath}

\begin{document}

\preprint{}

\title[Specific Heat of Mg$^{11}$B$_2$]{Specific Heat of Mg$^{11}$B$_2$}

\author{F. Bouquet}
\affiliation{Lawrence Berkeley National Laboratory and Department of Chemistry,
University of California, Berkeley, CA 94720}

\author{R. A. Fisher}
\affiliation{Lawrence Berkeley National Laboratory and Department of Chemistry,
University of California, Berkeley, CA 94720}

\author{N. E. Phillips}
\affiliation{Lawrence Berkeley National Laboratory and Department of Chemistry,
University of California, Berkeley, CA 94720}

\author{D. G. Hinks}
\affiliation{Materials Science Division, Argonne National Laboratory, Argonne, IL 60439}

\author{J. D. Jorgensen}
\affiliation{Materials Science Division, Argonne National Laboratory, Argonne, IL 60439}

\date{\today}

\begin{abstract}
Measurements of the specific heat of Mg$^{11}$B$_2$,
from 1 to 50 K,  in magnetic fields to 9 T, give the Debye
temperature, $\Theta$ = 1050 K, the coefficient of the normal--state electron
contribution, $\gamma_n$ = 2.6 mJ mol$^{-1}$ K$^{-2}$, and a
discontinuity in the zero--field specific heat of 133 mJ mol$^{-1}$ K$^{-1}$
at $T_c$ = 38.7 K. The estimated value of the electron-phonon
coupling parameter, $\lambda$ = 0.62, could account for the
observed $T_c$ only if the important phonon frequencies are
unusually high relative to $\Theta$. At low $T$, there is
a strongly field--dependent feature that suggests the
existence of a second energy gap, about four times
smaller than the major gap.
\end{abstract}

\maketitle

The superconductivity of MgB$_2$, with $T_c$ = 39 K
\cite{Nagamatsu} and an isotope effect \cite{Budko,Hinks,Lawrie}
consistent with the phonon--mediated electron pairing of the BCS
theory \cite{Bardeen}, has reopened the question of the maximum
$T_c$ that can be produced by that mechanism
\cite{McMillan,Allen}.  It also raises the complementary
question:  What is the mechanism of the superconductivity of
MgB$_2$ itself?  In this Letter, we report measurements of the
heat capacity ($C$) of Mg$^{11}$B$_2$ that give information
relevant to the latter question:  Comparison of the coefficient
($\gamma_n$) of the normal--state electron contribution to $C$
with band--structure calculations \cite{Sun,An,Kortus} gives an
estimate of the electron--phonon coupling parameter ($\lambda$).
The value of $\lambda$ suggests moderately strong coupling, but
whether it can account for the high value of $T_c$ on the basis
of phonon coupling alone depends on the values of other
parameters that are not yet determined. In the
superconducting state $C$
deviates from the BCS expression in a way that has no parallel
among known superconductors, and which suggests the presence of a
second, smaller gap in the electron density of states (EDOS),
which has a non--BCS dependence on $T$.

The Mg$^{11}$B$_2$ sample was a powder, prepared by reacting
$^{11}$B powder and Mg metal in a capped BN crucible at 850$^{o}$C
under a 50--bar argon atmosphere for 1.5 hours. Thermal contact
to the powder was achieved by mixing it with a small amount of
GE7031 varnish (a common low--$T$ thermal contact
agent with a known heat capacity) in a
thin--walled copper cup.  These extra contributions to the
addenda limited the precision of the data.  However, the
alternate method of providing thermal contact, sintering the
powder, can have adverse effects on the sample \cite{Franck} and
may in some cases account for differences between the results
reported here and those obtained in other measurements.  The
measurements of $C$ were made by a modified heat--pulse
technique, 1--32 K, and by a continuous--heating technique,
29--50 K. Measurements in magnetic field ($H$) were made on
the field--cooled (FC) sample after applying the field at $T \geq
60$ K.  For $H$ = 1 T, for which the field penetration is about
50\% of that in the normal state, measurements made after
applying the field at 1 K were indistinguishable from the FC
results, suggesting that equilibrium flux distributions were
attained.

Below 2 K, there is an $H$--dependent hyperfine contribution to
$C$. There are also several $H$--independent features in $C$,
including an ``upturn'' below 2 K, that are probably associated
with small amounts of impurity phases (or possibly with the GE
varnish, although past experience makes that seem unlikely).
Partly for that reason, most of the interpretation of the results
is based on an analysis of the differences, $C(H) - C(9 \text{
T})$, in which the $H$--independent extraneous contributions and
most of the contributions of the addenda, including the varnish,
cancel. $C(H) - C(9 \text{ T})$ was calculated after the data
were corrected for the hyperfine contributions and a small
$H$--dependent part of the heat capacity of the sample holder.
Since $C$ is the sum of an $H$--dependent electron contribution
($C_e$) and an $H$--independent lattice contribution ($C_l$), an
analysis of $C(H) - C(9 \text{ T})$ also has the advantage that
$C_l$ cancels, leaving $C_e(H)$, the contribution of greater
interest. In the normal state $C_e(H) = \gamma_n T$, independent
of $H$; in the mixed state $C_e(H)$ includes a $T$--proportional
term, $\gamma(H)T$, and $H$--dependent terms; in the
superconducting state, $C_e(0) = C_{es}$

The upper critical field ($H_{c2}$) of MgB$_2$ is approximately
linear in $T$ with $H_{c2}(0) \sim 16$ T \cite{Budko2}. This is
reasonably consistent with the $C$ measurements, for which the
onset of the transition to the mixed state, at $H = H_{c2}(T)$,
is marked by the deviations of $C(H) - C(9 \text{ T})$ from zero
(see Fig 1(a)). It leads to the expectation that for 9 T the
sample would be in the normal state for $T \geq 20$ K. The data
actually plotted in Fig. 1(a) are the differences, $[C(H) - C(9
\text{ T})]/T$, but with the scale shifted by  $\gamma$(9 T).
(They are essentially point--to--point differences, with no
smoothing of the 9--T data, which increases the scatter.) The
values of $\gamma(H)$ have been determined by fitting the
low--$T$, mixed-- and superconducting--state data with $[C(H) -
C(9 \text{ T})] = [\gamma(H) - \gamma(9 \text{ T})]T + a
\exp(-b/T)$, where $a$ and $b$ are $H$ dependent. Consistent with
the $H = 0$ data, $\gamma(0)$ was taken as zero, fixing the
values of $\gamma(H)$ for all $H$. Figure 2 shows $\gamma(H)$ vs
$H$, and the extrapolation to $H_{c2}(0) = 16$ T to obtain
$\gamma_n = 2.6$ mJ mol$^{-1}$ K$^{-2}$. Although the
extrapolation is somewhat arbitrary, the very small differences
between $\gamma$(5 T), $\gamma$(7 T), and $\gamma$(9 T) suggest
that it gives $\gamma_n$ to within $\sim 0.1$ mJ mol$^{-1}$
K$^{-2}$.  If the sample were normal at all temperatures in 9 T
the quantity shown in Fig. 1(a) would be exactly $C_e(H)/T$. That
is not the case, but the differences are small, as shown by the
small differences between $C$(5 T), $C$(7 T), and $C$(9 T).
Quantitatively, $C_e(H)/T$ is underestimated by the amount
$\gamma_n - C_e(9 \text{ T})/T$. For $T \leq 10$ K, $\gamma_n -
C_e(9 \text{ T})/T \approx \gamma_n - \gamma(9 \text{ T}) \approx
0.08$ mJ mol$^{-1}$ K$^{-2}$; for $10 \leq T \leq 20$ K, where
there must be a small broad anomaly in $C$(9 T), the
underestimate is smaller and $T$--dependent.

The transition to the superconducting state, shown in Fig. 1(c),
with an entropy--conserving construction that gives $T_c = 38.7$
K and $\Delta C(T_c) = 133$ mJ mol$^{-1}$ K$^{-1}$, is relatively
sharp, with a width $\sim$ 2 K. In this temperature interval the
sample is in the normal state for $H$ = 9 T, and addition of
$\gamma_n = 2.6$ mJ mol$^{-1}$ K$^{-2}$ to the quantity plotted
would give the electron contribution to $C$ through the
transition. The effect of $H$ in broadening the transition (to
the mixed state), as expected for measurements on a powder with
an anisotropic $H_{c2}$ \cite{An,deLima}, is evident in Fig. 1(a).

The thermodynamic consistency of the data, including in
particular the very unusual $T$ dependence of $C_{es}(0)$, can be
tested by calculating the difference in entropy ($S$)  between 0
and $T_c$ for different fields.  The result of such a test is
shown in Fig. 3(a) where the entropies obtained by integrating the
plotted points are compared with $\gamma(9 \text{ T})T$, which
represents the 9 T data.  At 40 K, the entropies for all $H$ are
within $\pm$ 2\% of the same value.  The result of a second
integration of the entropies to obtain free energy differences
and the thermodynamic critical field ($H_c$) is shown in Fig.
3(b).

Fitting
the 9--T data for $20 \leq T \leq 50$ K with $C(9 \text{ T}) =
\gamma_n T + C_l$, where $C_l = B_3 T^3 + B_5 T^5$, gave
$B_3 = 5.1\times10^{-3}$ mJ mol$^{-1}$ K$^{-4}$ and $B_5 =
2.5\times10^{-6}$ mJ mol$^{-1}$ K$^{-6}$.  The Debye temperature
($\Theta$), calculated following the usual convention of using the
value of $B_3$  per g atom, is $1050 \pm
50$ K.

Fig. 1(a) includes a comparison of the experimental $C_{es}$ with
that for a BCS superconductor with $\gamma_n = 2.6$ mJ mol$^{-1}$
K$^{-2}$ and $T_c$ = 38.7 K.  For $T \geq 27$ K, $C_{es}$ is
approximately parallel to the BCS curve; at lower $T$ it rises
above the BCS curve and then decreases to zero as $a \exp(-b/T)$,
but with values of $a$ and $b$ very different from those of the
BCS superconductor. It seems unlikely that this behavior could be
understood on the basis of gap anisotropy similar to that known in
other superconductors. Qualitatively, it gives the appearance of
a transition to the superconducting state in two stages:  the
first, a partial transition at $T_c$ that leaves a ``residual''
$\gamma$ (the extrapolation of $C_{es}$ to $T = 0$ from above 12
K gives $\sim 1$ mJ mol$^{-1}$ K$^{-2}$, but it should be
regarded as an overestimate because it does not allow for the
full entropy associated with the second stage); a second stage
that is associated with a second, smaller, energy gap, which
decreases in amplitude  in the vicinity of 10 K.  Taking the
exponential decrease in $C_{es}$ as a manifestation of a
BCS--like transition, and comparing the parameters $a$ and $b$
with BCS theory, gives $T_c$ = 11 K and $\gamma = 0.74$ mJ
mol$^{-1}$ K$^{-2}$.  This interpretation of $C_{es}$ is, to some
degree, understandable on the basis of theoretical
considerations:  It has been suggested that the gaps may be
different on two parts of the Fermi surface \cite{Liu}, and,
depending on the strength of the electron--phonon coupling
between, and on, the two parts, the amplitude of the small gap can
show a relatively abrupt decrease at a temperature well below
$T_c$.  Although both gaps open at $T_c$, that feature in the
small gap could produce the observed feature in $C_{es}$ in the
8--12 K region. The existence of two gaps on the Fermi surface is
also consistent with   scanning tunneling spectroscopy, which has
shown both a flat--bottomed BCS--like gap at low $T$, but with a
small amplitude, $\sim 2$ meV, corresponding to a BCS $T_c$ of
$\sim 13$ K \cite{Rubio}, and a V--shaped gap with an amplitude
of $\sim 5.2$ meV, corresponding to a BCS $T_c$ of 35 K
\cite{Karapetrov}, more compatible with the observed $T_c$.
Despite the difficulty of explaining why and how different groups
measure different gaps (see also \cite{Schmidt} and
\cite{Sharoni}), it is striking how well these two gaps would
account for $C_{es}$:  At low $T$, where the thermal excitations
are too weak to overcome the larger gap, the exponential behavior
of $C_{es}$ is consistent with the smaller gap; the opening of
the large gap at $T_c$ explains the large $\Delta C(T_c)$.
Moreover, between 20 K and $T_c$, $C_{es}/T$ has a linear
behavior (see Fig. 1(a)), which is consistent with a V--shaped
gap.  (Similar $T^2$ behavior has been seen in heavy--fermion
superconductors \cite{Fisher}.)

Anisotropy in $H_{c2}$ cannot explain the dramatic increase in
$\gamma(H)$ at low $H$ shown in Fig. 2. The dashed curve is a
calculation using the effective--mass model  with an anisotropy
of 10, which is already greater than reported values \cite{An,deLima},
but $\gamma(H)$ cannot be fitted with {\em any} value of the
anisotropy. The rapid increase in $\gamma(H)$ at low $H$
reflects the difference between $C_e(0.5 \text{ T})$ and $C_e(0)$
below 10 K, and is related to the existence of the second gap.

An average over the Fermi surface of the electron--phonon coupling
parameter can be estimated by comparing $\gamma_n$ with
band--structure calculations of the ``bare'' EDOS at the Fermi
surface ($N(0)$) using the relation $\gamma_n = (1/3) \pi^2
k_B^2N(0)(1+\lambda)$. $N(0)$ has been reported as 0.68, 0.71, and
0.72 states eV$^{-1}$ unit cell$^{-1}$ (Refs. \cite{Sun},
\cite{An}, and \cite{Kortus} respectively) giving $\lambda =
0.62$, 0.56, and 0.53.
Theoretically calculated values of $\lambda$ are 0.68 \cite{Sun},
and $\sim 1$ \cite{Kortus}. In the Mc Millan relation \cite{McMillan},
$T_c$ is related to $\lambda$, $\Theta$,
and the electron--electron repulsion ($\mu^*$) by
\begin{equation}
T_c = (\Theta /1.45)
\exp\{-1.04(1+\lambda)/[\lambda-\mu^*(1+0.62\lambda)]\},
\end{equation}
with $\mu^*$ frequently taken to  be $\sim$ 0.1
\cite{McMillan,Allen}. With $\lambda = 0.62$, the highest of the
values derived from $\gamma_n$ and $N(0)$, and $\Theta = 1050$ K
Eq. (1) gives $T_c = 22$ K, too low by almost a factor two. With
these values of $\lambda$ and $\Theta$, $T_c = 39$ K would
require $\mu^* = 0.030$, an unusually low value. However, in this
expression $\Theta$ represents a relatively crude estimate of the
phonon frequencies that are important in the electron pairing. As
derived from the coefficient of the $T^3$ term in $C_l$, $\Theta$
is really a measure of the frequencies of the low--frequency
acoustic phonons, which may not be particularly relevant to the
pairing of the electrons. In the Debye model, $\Theta$ is also
the cut--off frequency, but real phonon spectra often extend to
significantly higher frequencies. In the Allen and Dynes version
of the theory \cite{Allen}, $\Theta/1.45$ is replaced by
$\omega_{log}/1.20$, where $\omega_{log}$ is a moment of the
phonon frequencies in which they are weighted by the
electron--phonon matrix elements. More detailed calculations that
take into account relevant features of the phonon spectrum and
electron--phonon scattering may give a value of the
pre--exponential factor in Eq. (1) that accounts for the observed
$T_c$ with a physically plausible value of $\mu^*$, but until
they are available, the question of whether the electron pairing
in MgB$_2$ is phonon mediated would seem to remain open.

The parameters $(H_c(0))^2/\gamma_nT_c^2$ and $\Delta
C(T_c)/\gamma_nT_c$  measure  the
strength of the electron pairing \cite{Padamasee}.
In the BCS, weak--coupling
limit, their values are 5.95 and 1.43,
respectively. There are a number of ``strong--coupled''
superconductors for which these parameters are greater than the
BCS values, but relatively few for which they are smaller
\cite{Padamasee}. For Mg$^{11}$B$_2$ they are
unusually small, 5.46 and 1.32, which, in contrast with the
value of $\lambda$, suggests extreme weak coupling. This
indication of weak coupling, based on bulk properties, is
in qualitative agreement with tunneling results that may have
been influenced by surface effects \cite{Karapetrov}.

In general, there are more differences than similarities among
the specific heat measurements on MgB$_2$
\cite{Budko,Wang,Kremer,Marcenat,Walti}. However, in several
important respects, including the low--$T$ behavior of $C_{es}$,
the results of the Geneva group \cite{Wang} are qualitatively
similar to those reported here, and the differences that do occur
are readily understood in terms of sample dependence. A
comparison with their results, which were obtained by different
experimental techniques on sintered sample of commercial
material, attests the qualitative validity of the major features
reported both here and in Ref. \cite{Wang}:
The value of $\gamma_n$, 2.7 mJ mol$^{-1}$ K$^{-2}$, and the test of thermodynamic consistency in Ref. \cite{Wang} are similar and of comparable accuracy to those reported here, but limited by paramagnetic Fe impurities in Ref. \cite{Wang} and the precision of the data here. Although the superconducting--state entropies are essentially identical at $T_c$, the specific heat anomaly at $T_c$ in Ref. \cite{Wang} is broader, leading to a substantial underestimate of $\Delta C(T_c)$. Most importantly, the
$T$--dependence of $C_{es}$ in the 4--15 K range, which shows the
presence of a second gap (or extreme an unusual anisotropy), is
essentially the same in both cases. However, for $T \leq 4$ K,
$C_{es}$ was obscured by the contribution of paramagnetic
impurities and the limiting $T \rightarrow0$ dependence was taken
as approximatively $T^2$ \cite{Wang}, which would be expected for
line nodes, rather than the exponential dependence reported here.

Several other values of $\Delta C(T_c)$, all for sintered samples
and all lower than that reported here, have been given in other
reports \cite{Budko,Kremer,Marcenat,Walti}. Values of $\gamma_n$ that range from 1.1 to
5.5 mJ mol$^{-1}$ K$^{-2}$, based on different \cite{Kremer,Walti} or
unspecified \cite{Budko} analysis of experimental data have also been
reported, but the very similar values reported here and in Ref. \cite{Wang},
 are supported by the thermodynamic
consistency of the data

We have benefited from useful discussions with J. M. An, M. L.
Cohen, G. W. Crabtree, J. P. Franck, R. A. Klemm, C. Marcenat, I.
I. Mazin, and W. E. Pickett. The work at LBNL
was supported by the Director, Office of
Basic Energy Sciences, Materials Sciences Division of the U. S.
DOE under Contract No. DE--AC03--76SF00098.  The
work at ANL was supported by the U. S.
DOE, BS-MS under Contract No.
W--31--109--ENG--38.

\newpage

\begin{figure}
\caption{(a)  $[C(H) - C(9 \text{ T})]/T$. In (a) and (b) the
scale has been shifted by $\gamma(9 \text{ T})$ to give an
approximation to $C_e(H)/T$ (see text). In (a) the dashed curve
is a polynomial extrapolation of the 12--20 K, $H = 0$ data to $T
= 0$; the horizontal line represents $\gamma(9 \text{ T})$. In
(b) the low--$T$ 5-- and 7--T data are shown on an expanded scale
with solid curves representing fits described in the text; the
horizontal solid and dashed lines represent $\gamma(9 \text{ T})$
and $\gamma_n$, respectively. In (c) the solid lines represent an
entropy conserving construction.  The error bars are $\pm$ 0.1\%
$C_{total}/T$.}
\label{fig1}
\end{figure}

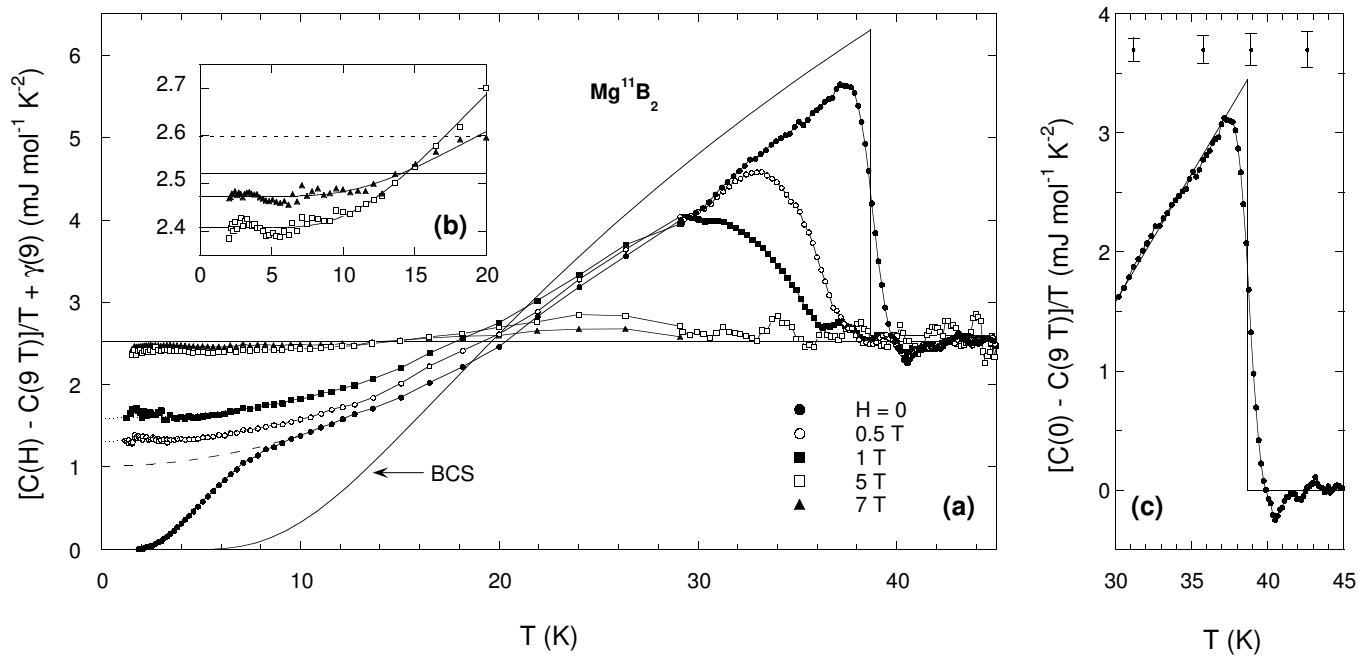
\begin{figure}
\caption{$\gamma$ as a function of $H$. The  solid curve is a
guide to the eye and an extrapolation to $H_{c2}(0)$. The dashed
curve is a ``fit'' with an $H_{c2}$ anisotropy of 10.}
\label{fig2}
\end{figure}

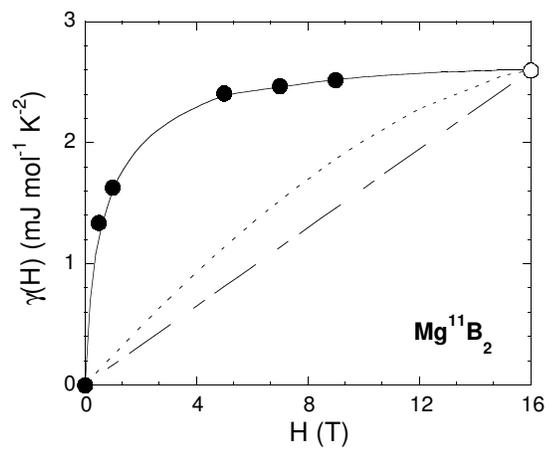
\begin{figure}
\caption{(a) Entropies as functions of $T$ for different $H$, with
$H$ increasing from the lowest to highest curve.  (b)
Thermodynamic critical field, compared with a BCS curve for the
derived values $\gamma_n$ and $T_c$.}
\label{fig3}
\end{figure}

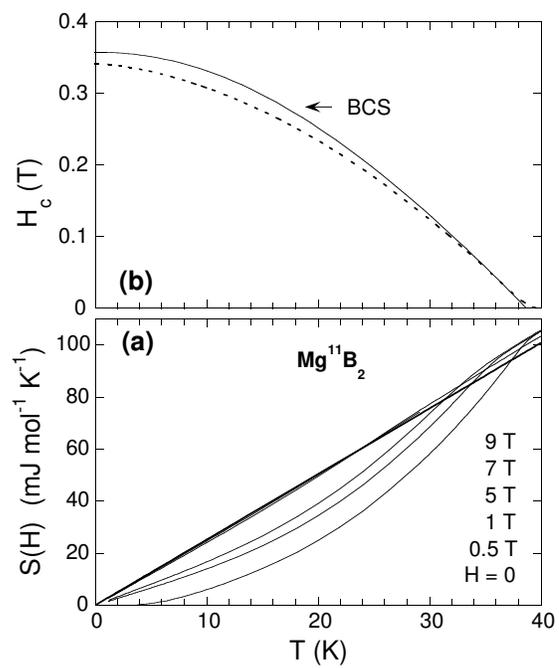
\end{document}